\documentstyle[prl,aps,epsfig]{revtex}
\begin{document}
\twocolumn[\hsize\textwidth\columnwidth\hsize\csname
@twocolumnfalse\endcsname

\draft

\title{Fluctuation - induced nucleation and dynamics of the kinks on dislocation.
Soliton and oscillation regimes in 2D Frenkel-Kontorova model. }

\author{
Yu. N. Gornostyrev$^{1}$, M.I.Katsnelson$^{1}$, A.V.Kravtsov$^{1}$,
and  A.V.Trefilov$^{2}$
}
\address{
\(^1\)  Institute of Metal Physics, Ekaterinburg 620219, Russia.\\
\(^2\)   Russian Research Centre "Kurchatov Institute",
 Moscow 123182, Russia. 
}

\maketitle

\begin{abstract}
Numerical simulation of the dislocation motion in 2D Frenkel - Kontorova
(FK) model in the thermostat shows an unusual dynamical behavior. It appears
that ''kink'' regime of dislocation gliding takes place in a certain region
of parameters of the model but, in disagreement with the common views about
the dislocation motion under plastic deformation condition, the kinks appear
to be similar to sine-Gordon solitons despite the discreteness of the
lattice, damping and thermal fluctuations. At high enough stresses and
temperatures the motion of the dislocation is accompanied by its
oscillations rather than kink nucleation.
\end{abstract}

\pacs{
  05.45,        61.72.Lk,     61.72.Bb,   }

\vskip2pc] 

\section{Introduction}

The concept of solitons appeared to be very fruitful in various fields of
condensed matter physics such as magnetism, charge-density waves, electronic
structure of polymers, etc. \cite{Maki}. Soliton is defined as a nonlinear
wave in continuum media with particle-like properties, i.e. conserving its
form both at the propagation and at the collisions with other solitons.
Generally speaking, the literal transfer of the results of soliton theory to
the case of crystal lattice is impossible since the taking into account the
discreteness of the lattice may change drastically the character of
solutions of nonlinear equations. In particular, the discreteness eliminates
the Goldstone mode leading to the pinning of solitons, create the new
classes of localized nonlinear oscillatory solutions (breathers), induces
the coupling of ''soliton'' and ''phonon'' modes and so on \cite{Maki}, \cite
{Flach}.

Apart from this, the effects of dissipation, thermal fluctuations etc. are
usually important in real situations. There has been recently a growing
interest in non-linear effects in the crystal lattice dynamics particularly
in the phenomena responsible for plastic deformation (see, e.g., \cite
{Nikitenko97}). According to common opinions \cite{Hirth}, the gliding of
dislocations in crystals with sufficiently high Peierls barrier (in
particular, in covalent crystals and semiconductors) is by thermofluctuation
creation of kink pairs (kink/antikink). In the course of their motion in the
opposite sides under the action of external stress the segment of
dislocation between the kink and antikink glides into the neighbor Peierls
barrier valley. However, while the existence of such kinks is confirmed by
direct electronmicroscopic examinations \cite{Alexander}, the information on
the character of their motion is indirect , i.e. it is obtained from the
comparison of results of particular theoretical models \cite{Seeger},\cite
{Lothe},\cite{Hirsch} with the measurements of dislocation mobility. The
common feature of all these models is that the kink is considered as a
solitary wave in the Frenkel-Kontorova (FK) model. That is, it is supposed
that its shape would persist in the motion \cite{Kosevich},\cite{Pawelek}.
This assumption, strictly speaking, has not been proven since the analytical
solutions of soliton type are only available for the ideal continuum model,
and the question on their existence in more realistic discrete case is still
open.

In the continuum limit the 1D FK model is reduced to the well - known
sine-Gordon model \cite{Aubry} which is completely integrable. This means
that any initial disturbance of the finite amplitude in this model breaks up
into the succession of kinks (solitons), antikinks and breathers (soliton
/antisoliton coupled states). In this case the soliton/antisoliton collision
may occur by the mechanism of elastic (i.e. without change in the shape)
scattering or be accompanied by breather formation. It is unknown to what
extent such a behavior might be applied to the discrete case. The existing
phenomenological model of kink motion along a dislocation \cite{Hirth}, \cite
{Alexander} are based on a quite contrary supposition of an unavoidable
kink-antikink annihilation in the collision. This statement bases on the
hypothesis that such factors as the discreteness of crystal lattice, damping
(viscosity) and thermal fluctuations, so important in the actual situation,
would destroy the features associated with the complete integrability. Of
course, this hypothesis can be either confirmed or refuted only by means of
particular numerical calculations which, however, did not give any definite
answer to this questions up to now \cite{Srolovitz}. It should be stressed
also that applications of results concerning soliton motion in 1D models to
the kinks are doubtful even in the continuum limit since the kinks are
''solitons on soliton'', i.e. nonlinear excitations of 2D or 3D character.
We do not know any analytical results about such excitations in
non-one-dimensional continuum models and therefore computer simulation is
the only way to investigate them.

Ref. \cite{Srolovitz} was the first attempt to do this but the nucleation
and propagation of kinks were not observed in that calculation. Only
recently opportunities appeared of computer simulations of the dislocation
motion in realistic 3D case. Rather complicated picture of the dislocation
motion in Al has been demonstrated in \cite{MLi}. However, since even more
simple 2D situation is not understood now, it is worthwhile to investigate
systematically processes of nucleation, propagation and destruction of kinks
for a model case.

The present work is devoted to computer simulations of that processes in the
framework of 2D FK model. We show that, for certain values of the model
parameters, the dislocation gliding is really accompanied by the nucleation
and motion of kink-antikink pairs but the dynamics of the latter can be
soliton-like. At the same time, for another region of the model parameters
well-defined kinks are not observed at all. It means that basic assumptions
of conventional phenomenological models of dislocation gliding are at least
not evident and the question about their applicability remains.

\section{Formulation of the model and simulation method}

Consider the common two-dimensional generalization of FK model (2D FK). To
take into account the effects of damping and interaction with the thermostat
we solve numerically (as well as in \cite{Srolovitz}) the Langevin equations
of the following form 
$$
\stackrel{\cdot \cdot }{u}_{ni}=-\frac{\partial V}{\partial u_{ni}}-\gamma 
\stackrel{\cdot }{u}_{ni}+\xi _{ni}(t)+f_i,\eqno(1) 
$$
where ${\bf u}_n$ is the vector of $n$-th atom displacement, $i=(x,y)$ are
the Cartesian coordinates, $\gamma $ is the damping factor (the atom masses
being equal to 1), $f_i$ is the constant external force, $\xi _{ni}(t)$ is
the Gauss random force ($<\xi _{ni}(t)>=0$) with the correlator 
$$
<\xi _{ni}(t)\xi _{n^{\prime }i^{\prime }}(t^{\prime })>=2\gamma T\delta
_{nn^{\prime }}\delta _{ii^{\prime }}\delta (t-t^{\prime }),\eqno(2) 
$$

As is known \cite{VanKampen} such a choice of random force guarantees the
Gibbs distribution to be established in the equilibrium at temperature $T$.
The potential energy is chosen as 
$$
V=\frac K2\sum\limits_{<n,m>}({\bf u}_n-{\bf u}_m)^2+P\sum\limits_n\sum%
\limits_{{\bf g}}(1-\cos {({\bf g\cdot u}}_n)),\eqno(3) 
$$
where $<n,m>$ are the pairs of nearest neighbors in the lattice, {\bf g} is
three least vectors of reciprocal lattice (we will consider the hexagonal
lattice with a parameter $a=1$, then ${\bf g}_1=\frac{4\pi }{\sqrt{3}}(-1,0),%
{\bf g}_{2,3}=\frac{4\pi }{\sqrt{3}}(\frac 12,\pm \frac{\sqrt{3}}2)$). The
first term in Eq.(3) describes the interaction of atoms in the harmonic
approximation, and the second one is the simplest approximation
corresponding to the sine law for the returning force in 1D FK model. Note
that, unlike the present work, Ref. \cite{Srolovitz} considered the square
lattice which, in the nearest neighbor approximation, has a certain
pathology (the zero shear modulus). In contrast with this, hexagonal lattice
in the nearest--neighbor approximation represents rather general isotropic
2D situation. Thus, Eq. (3) can be considered as the most natural
generalization of original 1D FK model for 2D case. Depending on specific
interatomic interactions and chemical bonds, the substrate potential may be
changed in comparison with purely cosine form in Eq. (3). Calculating the
substrate potentials for a series of fcc metals \cite{Ir} we found that Eq.
(3) may be considered as an adequate enough for substances with almost
purely central pairwise interactions (without strong covalency or electron
density dependence). In principle, it is not difficult to take into account
next terms of Fourier expansion in Eq. (3) but in this work we restrict
ourselves only by the consideration of the simplest case .

Initially lattice displacements of atoms ${\bf u}_n(0)$ corresponding to the
screw dislocation with Burgers vector {\bf b}=(0,1) and the axis parallel
OY, were set. The displacements for $t=0$ were chosen in accordance with the
solution of 1D FK as \cite{Hirth} 
$$
u_{ny}=\frac 2\pi \left( \frac \pi 2-tan^{-1}\left[ exp\left( -\frac{x_n}%
\lambda \right) \right] \right) ,\;u_{nx}=0\eqno(4) 
$$
where $\lambda =\frac 1\pi \sqrt{\frac KP}$. Eqs.(1) were then solved for $%
f=0$, $T=0$ and the ''actual'' equilibrium distribution of displacement
field in the absence of external force was determined. After that external
force $f_y$ ($f_x=0$) and temperature $T$ was introduced, and the evolution
of displacement field was studied. The lattice was specified the form of
40x40 cluster of atoms with periodic boundary conditions. Generally
speaking, the latter may be dangerous in the simulations of dislocations in
realistic models with inverse distance decay of deformation fields and
therefore with strong interactions with ''mirror'' dislocations. However,
the deformation in FK model decay exponentially and long-range part of the
dislocation fields is not taken into account.

In order to obtain a reliable information about the evolution of the system
investigated, it is necessary to use the so-called ''strong'' (mean square)
methods for integrating stochastic differential equations \cite{Milshtein},%
\cite{GKTT}. The simpler and more commonly used weak methods (in particular,
those used in \cite{Srolovitz}) are actually intended only for the
calculation of averaged characteristics such as spectral density and are not
appropriate for studying individual trajectories of the system in the
thermostat. We used the method proposed in \cite{Milshtein2} and intended
for solving the set of stochastic differential equations (the approach used
in \cite{GKTT} is only directly applicable for the solution of single
equation). Specific computational formulas and the details of the method are
presented in Appendix.

\section{The results of calculations}

As is seen from Eqs. (1)-(3) the character of solution is determined by four
independent dimensionless parameters $\widetilde{\gamma }=\gamma /\sqrt{K}$, 
$\widetilde{P}=P/K$, $\widetilde{T}=T/K$, and $\widetilde{f}=f_y/K$.
Dimensionless damping $\widetilde{\gamma }$ is of order of the ratio of
phonon damping to phonon frequencies and is typically of order of 10$^{-2}$
at room temperature and above \cite{VaksKT}. Apart from this, such choose of 
$\widetilde{\gamma }$ provides strong attenuation of phonon waves at the
lengths of order of the crystallite size in our simulations and eliminate
essentially artifacts of periodic boundary conditions. Since the phonon
damping in the classical region is linear in T (see e.g. \cite{VaksKT}) we
put $\widetilde{\gamma }_1/\widetilde{\gamma }_2=\widetilde{T}_1/\widetilde{T%
}_2$.

The value of the parameter $\widetilde{P}$ determines the transition from
continuum case to the discrete one. One can estimate from the expression for 
$\lambda $ in Eq. (4) that the width of the kink is of order of interatomic
distance at $\widetilde{P}\simeq 0.1.$ As it will be seen below it is
roughly the ''critical'' value for the change of a character of dislocation
motion. Therefore we investigate in detail the cases $\widetilde{P}=0.08$
and $\widetilde{P}=0.12.$ As for the parameter of external stress it has to
be compared with the Peierls stress $f_p.$ The latter is defined as the
minimal stress which leads to the motion of the dislocation as a whole at $%
T=0$. The results of our simulations show that this critical stress turns
out to be temperature-dependent decreasing with $T$ increase. For example,
for $\widetilde{P}=0.08$, $\widetilde{T}=0.005$ the dislocation moves as a
practically straight line already at $\widetilde{f}=0.012-0.016$, whereas
Peierls stress for that parameters $\widetilde{P}$ and $T=0.002$ was $%
0.030-0.035$. The accurate calculation of the Peierls barrier in computer
simulation is a complicated problem \cite{VitekChristian}. Since we are
interested here in the investigation of the gliding we restrict ourselves
only the case $f<f_p\left( T\right) .$

Fig. \ref{fig:displ} gives an example of the distribution of displacements 
around the moving dislocation. Kinks may be seen as inhomogeneities of the 
distribution along the dislocation axis. However, it is much more suitable to 
watch them displaying only dislocation line positions. The latter is defined by 
the condition $|u_y|=0.5b$. It corresponds to the center of dislocation since 
$|u_y|$ varies continuously from 0 to $b$, see Fig. \ref{fig:displ}.

Fig. \ref{fig:narrowkinks} demonstrates a typical (at least for the values
of temperature and damping under consideration) picture of dislocation
gliding for $\widetilde{P}=0.12.$ It shows ''soliton--like'' behavior of
kinks with their thermofluctuation nucleation, propagation without change of
their shapes and purely elastic scattering (mutual penetration) of kink and
antikink. To our knowledge, this is the first observation of such regime in
2D FK model with the interaction with thermostat.

Figs. \ref{fig:widekinks},\ref{fig:oscillation} display some pictures of the 
gliding for $\widetilde{P}=0.08.$ The 
character of the dislocation motion for this value of $\widetilde{P}$ is
much more complicated than for $\widetilde{P}=0.12$ . Depending on the
parameters of the model three different regimes of the dislocation gliding
are possible. First, kink nucleation similar to Fig. \ref{fig:narrowkinks} 
may take place. Second, kinks may appear after preliminary stage of 
''oscillation'' motion when the dislocation glides as a wriggled flexible band 
or, by another words, kinks arise as a result of evolution of the oscillations 
of the dislocation segment (Fig. \ref{fig:widekinks}). The most important 
feature here is the demonstration of a possibility of a third regime, namely, 
rather fast gliding without participation of kinks at all, only due to the 
oscillations of the dislocation line (Fig. \ref{fig:oscillation}). Sometimes 
the average dislocation velocity without participation of kinks is even higher 
(compare the motion of dislocation line in Fig. \ref{fig:widekinks} where kinks 
exist and in Fig. \ref{fig:oscillation} without kinks).
Since the oscillating dislocation does not lie in a single Peierls valley
the ''effective'' Peierls barrier for such dislocation segment turns out to
be lower than for straight one. Results obtained here shows that the regime
of ''wriggled flexible band'' occurs at large enough temperature and
external stress close to the Peierls stress. Similar gliding mechanism due
to oscillations has been observed recently in molecular dynamics simulations
of the 3D dislocation motion in Al \cite{MLi}. According to our simulations,
the transitions between regimes occur at the change of temperature and
damping, e.g., for $T=0.002$ three regimes mentioned above take place for $%
0.02\leq $ $\widetilde{\gamma }$ $\leq 0.03,$ $0.03\leq $ $\widetilde{\gamma 
}$ $\leq 0.04,$ and $\widetilde{\gamma }$ $\geq 0.04,$ correspondingly.
General trend is the change of kink regime by the oscillation one at the
temperature and/or damping increase. The value of external stress is less
important provided that it is lower than the Peierls stress for a given
temperature.

Another important feature of that regime is the existence of long-lived
coupled kink-antikink pairs similar to discrete breathers \cite{Flach}. An
example of such behavior is seen in the part of Fig. \ref{fig:widekinks} 
corresponding to $24\le t\le 34$. It is worthwhile to stress that we mean 
breather-like excitations on the ''soliton'' (dislocation) and not standard 
discrete breathers as more or less isotropic localized oscillating in time 
solution of non-linear equations. Similar breather-like excitations may be 
observed also for $\widetilde{P}=0.12$ but their lifetime is much shorter than 
for $\widetilde{P}=0.08.$ Note that the lifetime increases with 
$\widetilde{\gamma }$ increase for given values of other parameters of the model.

We never observed the kink-antikink annihilation in our simulations; if
kinks appear they always behave as solitons at the collisions. It is rather
unexpected in comparison with common views \cite{Hirth},\cite{Alexander},%
\cite{Aubry}. To clarify the situation we investigate kink-antikink
collisions generating kink-antikink pairs right in the initial
configuration. The results of the simulations shown in Fig. 
\ref{fig:anihilation} demonstrate that both penetration and annihilation are 
possible depending on the parameters of the model, the annihilation being take 
place at rather high values of the damping and temperature 
(Fig. \ref{fig:anihilation}b). However, it is that the region of the parameters 
when, instead of thermofluctuation nucleation of kinks, oscillation regime of 
the gliding takes place. Therefore we believe that the kink-antikink 
annihilation is not important process for the dislocation gliding under 
''normal'' conditions when the creation of kinks is of purely thermofluctuation 
nature. However, it may be important if kinks could create, e.g., by collisions 
of the dislocation with another defects, etc. This question needs further 
investigations

\section{Discussion and Conclusions}

Summing up the results obtained, it might be stated that the physical
picture of the dislocation gliding appeared to be much more complicated that
it is supposed in traditional models of dislocation motion in the Peierls
relief \cite{Hirth},\cite{Alexander}. First of all, we observe (for $%
\widetilde{P}=0.12$) thermofluctuation nucleation of kinks supposed in that
models but, in contrast with them, the character of motion of the kinks
appears to be purely ''solitonic''. It is nontrivial since for these values
of the parameters the widths of kinks are of order of the lattice constant
and the results of continuum models such as sine-Gordon model are definitely
not applicable. This implies the annihilation of kinks and antikinks is
unlikely in this regime, and their lifetime is anomalously long. It means
that under these conditions the thermodynamical equilibrium in the system of
kinks and antikinks does not practically establish, and their concentration
should be essentially higher than that predicted by the phenomenological
models. Note that in the special pulse regime of loading the anomalously
high concentration of kinks was observed in Si \cite{Nikitenko2} and Ge \cite
{Nikitenko97}. The explanation proposed in \cite{Nikitenko97} is based on
the essential effect of impurities. According to the results obtained in the
present work this explanation is not the only possible one and high
concentration of kinks may be intrinsic property of that materials.

For larger values of width of kinks ($\widetilde{P}=0.08$) the character of
dislocation motion is very complicated. In contrast with the case $%
\widetilde{P}=0.12$ the creation of kinks in this case is not typical
thermofluctuation nucleation but rather a result of more or less long
evolution of oscillations of the dislocation segment. Both kinks and
oscillations of dislocation segments play an important role in the gliding.
It is shown that the gliding with rather high velocity may be provided by
oscillations only without participation of kinks (''wriggled flexible band''
regime). More detailed investigation of these processes depending on the
temperature, damping, external stress and the shape of substrate potential
relief would allow to correct and improve existing phenomenological models
of plastic deformation.

In the conclusion it should be noted that the methods used here can be
applied for studying variable loading conditions with a time-dependent
external force (ultrasound, pulse loads). The question on the role of
impurities \cite{Nikitenko97} might be also studied be means of similar
simulation. At the same time, we believe that the results obtained may be of
more general interest than for physics of plasticity only since until now
rather incomplete information about soliton-like excitations in
two-dimensional discrete systems are available.

\section{Acknowledgments}

The authors are grateful to S.V.Tret'jakov for helpful discussion of methods
for numerical solution of stochastic differential equations. The work is
supported by Russian Basic Research Foundation, grant 98-02-16219.

\section{Appendix}

Let us represent Eq. (1) as a set of first -order equations in a form used
commonly in the theory of stochastic differential equations (SDE)

$$
\begin{array}{c}
du_{ni}=v_{ni}dt \\ 
dv_{ni}=(-\nabla _{ni}V( {\bf u })-\gamma v_{ni})dt+\kappa d\omega _{ni}(t)
\end{array}
\eqno(A1) 
$$
where $\kappa =\sqrt{2\gamma T}$, $d\omega _{ni}(t)$ is a standard Wiener
process. Introducing $4nm$ -dimensional vectors

$$
\overrightarrow{x}=\left( 
\begin{array}{c}
{\bf u} \\ 
{\bf v}
\end{array}
\right) , \ \ \ 
\overrightarrow{a}=\left( 
\begin{array}{c}
{\bf v} \\ 
-\nabla V({\bf u})-\gamma {\bf v}
\end{array}
\right) , 
$$
$$
\overrightarrow{\sigma _r}=\left( 
\begin{array}{c}
{\bf 0} \\ 
{\bf b}_r
\end{array}
\right) 
$$

$$
b_{ri}=\kappa \delta _{ri},\ \ \ r=1,...,q,\ \ \imath =1,...,q, \ \ \ q=2nm, 
$$

one can rewrite Eqs. (A1) in a common form as

$$
d\overrightarrow{x}=\overrightarrow{a}(t,\overrightarrow{x}%
)dt+\sum\limits_{r=1}^{q}\overrightarrow{\sigma _r}d\omega _r(t) \eqno(A2) 
$$

To integrate SDE's (A2) numerically we use the mean-square Runge-Kutta type
method of the 3/2- order which leads to the following approximate solution 
\cite{Milshtein2}

$$
\overrightarrow{x}_{k+1}=\overrightarrow{x}_k+\frac 16(\overrightarrow{K}_1+2%
\overrightarrow{K}_2+2\overrightarrow{K}_3+\overrightarrow{K}%
_4)+\sum\limits_{r=1}^q(\overrightarrow{\sigma }_rI_r)_k+ 
$$
$$
\sum\limits_{r=1}^q(\frac{d\overrightarrow{\sigma }_r}{dt}%
I_{0r})_k+\sum\limits_{r=1}^q(\Lambda _r\overrightarrow{a}I_{r0})_k+\frac{h^2%
}2L_2\overrightarrow{a}_k+ 
$$
$$
\sum\limits_{r=1}^q(\frac{d^2\overrightarrow{\sigma }_r}{dt^2}%
I_{00r})_k+\sum\limits_{r=1}^q(\Lambda _rL_1\overrightarrow{a}%
I_{r00})_k+\sum\limits_{r=1}^q(L_1\Lambda _r\overrightarrow{a}I_{0r0})_k. 
$$

Here $\overrightarrow{x}_k$ is the mean-square approximation of the vector $%
\overrightarrow{x}$ at the instant $t_k,$

$$
\overrightarrow{K}_1=h\overrightarrow{a}(t_k,\overrightarrow{x}_k), \ \ \ 
\overrightarrow{K}_2=h\overrightarrow{a}(t_k+h/2,\overrightarrow{x}_k+ 
\overrightarrow{K_1}/2), 
$$
$$
\overrightarrow{K}_3=h\overrightarrow{a}(t_k+h/2,\overrightarrow{x}_k+ 
\overrightarrow{K_2}/2), 
$$
$$
\overrightarrow{K}_4=h\overrightarrow{a}(t_{k+1}, \overrightarrow{x}_k+ 
\overrightarrow{K}_3) 
$$

$$
L_1=\frac \partial {\partial t}+\sum\limits_{i=1}^\rho a^i\frac \partial {%
\partial x^i},\ \ \ L _2=\frac 12\sum\limits_{r=1}^q\sum\limits_{i,j=1}^\rho
\sigma _r^i\sigma _r^j\frac{\partial ^2}{\partial x^i\partial x^j},
$$
$$
\Lambda _r=\sum\limits_{i=1}^\rho \sigma _r^i\frac \partial {\partial x^i},
\ \ \ 
I_r=h^{1/2}\zeta _r,\ \ \ I_{r0}=\frac{h^{3/2}}2\left( \frac{\eta _r}{\sqrt{3%
}}+\zeta _r\right) ,
$$
$$
I_{0r}=hI_r-I_{r0},\ \ \ I_{r00}=hI_{r0}-J_r,%
$$
$$
I_{0r0}=2J_r-hI_{r0},\ \ \ I_{00r}=\frac{h^2}2I_r-J_r,
$$
$$
J_r=h^{5/2}\left( \frac{\zeta _r}3+\frac{\eta _r}{4\sqrt{3}}+\frac{\xi _r}{12%
\sqrt{5}}\right) ,
$$
$h=t_{k+1}-t_k;$ $\rho =4nm$ is the dimension of $\overrightarrow{x}$
vector; $\zeta _r,\eta _r$ and $\xi _r$ are independent normally distributed
random variables with zero mean value and dispersion equal to unity.

Generally speaking, this method is ''$h^{3/2}-$method'', i.e. it provides
the accuracy of order of $h^{3/2}$ for $\overrightarrow{x}_k$ for every
realization of stochastic process. However it may be proven \cite{Milshtein2}
that for the case of small enough noise it is even more accurate (''$h^4$
method''). We have chosen in our simulations $h=0.005;$ in that case the
accuracy of the calculations for trajectories as well as of Gibbs
distribution at the equilibrium appeared to be restricted only by
Monte-Carlo inaccuracy.


\section*{Captions to figures}

\begin{figure}[!htb]
\caption{{\small
A typical picture of the displacement distribution for the moving
dislocation at $\widetilde{P}=0.08, \widetilde{T}=2\cdot 10^{-3},
\widetilde{\gamma }=0.02, \widetilde{f}=-0.008$. 
Coordinates $x,y$ are in units of $a,$ $u_y$ (vertical axis) is in units of 
Burgers vector. 
}}
\label{fig:displ}
\end{figure}

\begin{figure}[!htb]
\caption{{\small
The configuration of the dislocation line is shown at different times $t=$ 10, 
13, 14, 15, 16, 18, 20, 22, 25, 27, 29, 31, 33, 35, for $\widetilde{P}
=0.12, \widetilde{T}=2\cdot 10^{-3}, \widetilde{\gamma }=0.02, 
\widetilde{f}=-0.151$. The pictures correspond to the consequent events from left to
right, the second row is the continuation of the first one. Dashed line
shows the initial position of the dislocation. Circles indicate position
of the atoms.
}}
\label{fig:narrowkinks}
\end{figure}

\begin{figure}[!htb]
\caption{{\small
As in  Fig. 2, $t=$ 10, 13, 16, 18, 20, 22, 24, 26, 28, 30, 32, 34, 37, 40,
for $\widetilde P =0.08, \widetilde T = 2\cdot 10^{-3}$, $\widetilde {\gamma}
=0.035$, $\widetilde f = -0.008$.
}}
\label{fig:widekinks}
\end{figure} 

\begin{figure}[!htb]
\caption{{\small
As in Fig. 2, $t=$ 7, 10, 13, 16, 19, 22, 25, 28, 31, 34, 37, 40, 43, 46, for 
$\widetilde{P}=0.08, \widetilde{T}=2\cdot 10^{-3}, \widetilde{\gamma }=0.045,
\widetilde{f}=-0.01$. 
}}
\label{fig:oscillation}
\end{figure} 

\begin{figure}[!htb]
\caption{{\small
The dynamics of the kink-antikink pair created in the initial
configurations. The dislocation line at different times $t=$ 4.5, 6, 7.5, 9,
10.5, 12, 13.5, 15 is shown. Here $\widetilde{P}=0.08, \widetilde{T}=2\cdot
10^{-3}$and (a) $\widetilde{\gamma }=0.02, \widetilde{f}=-0.030$ (penetration
regime) (b) $\widetilde{\gamma }=0.03, \widetilde{f}=-0.040$ (annihilation
regime).
}}
\label{fig:anihilation}
\end{figure} 

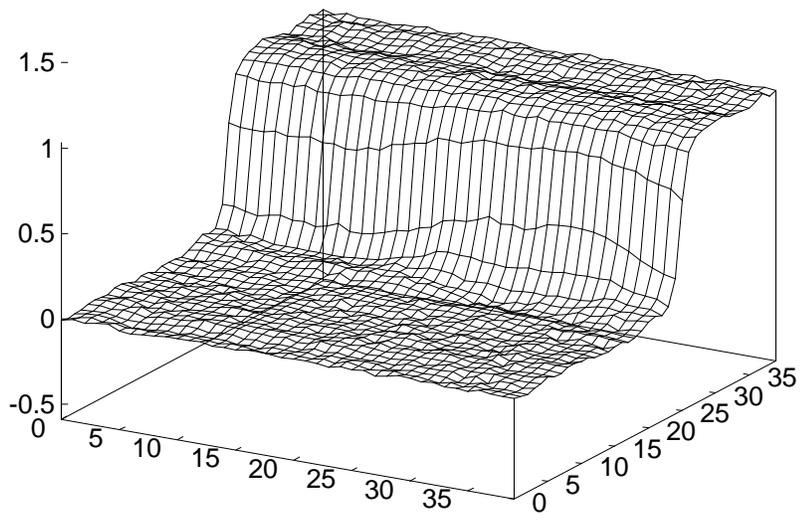
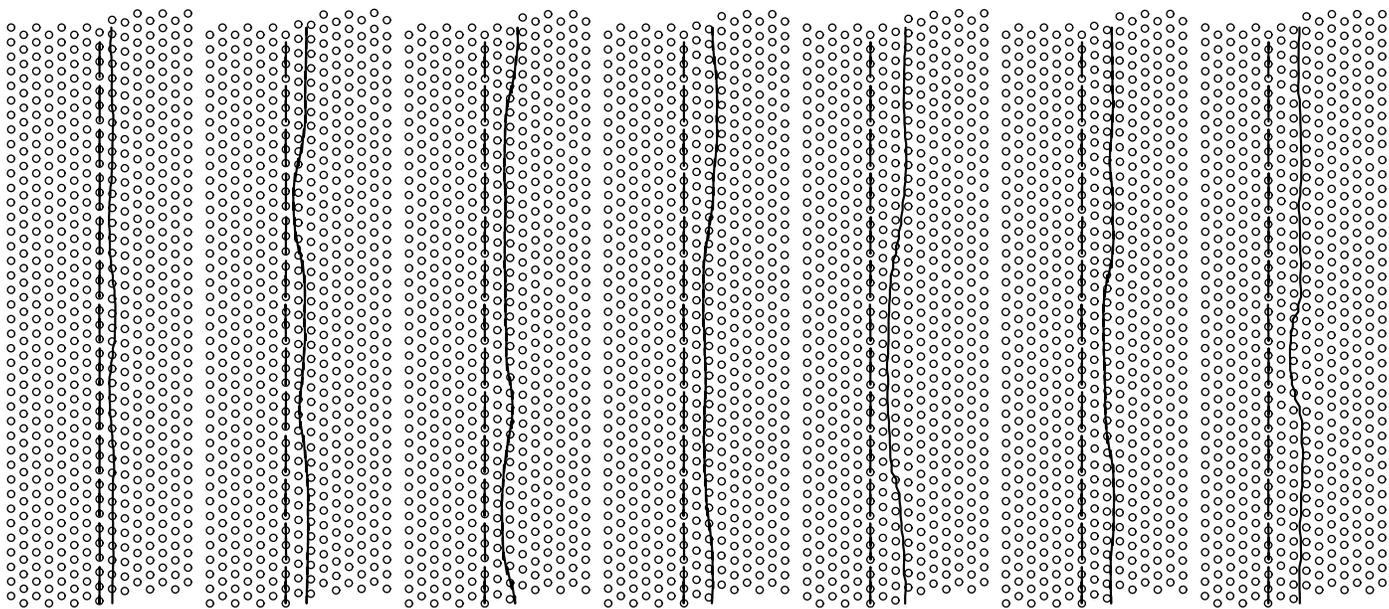
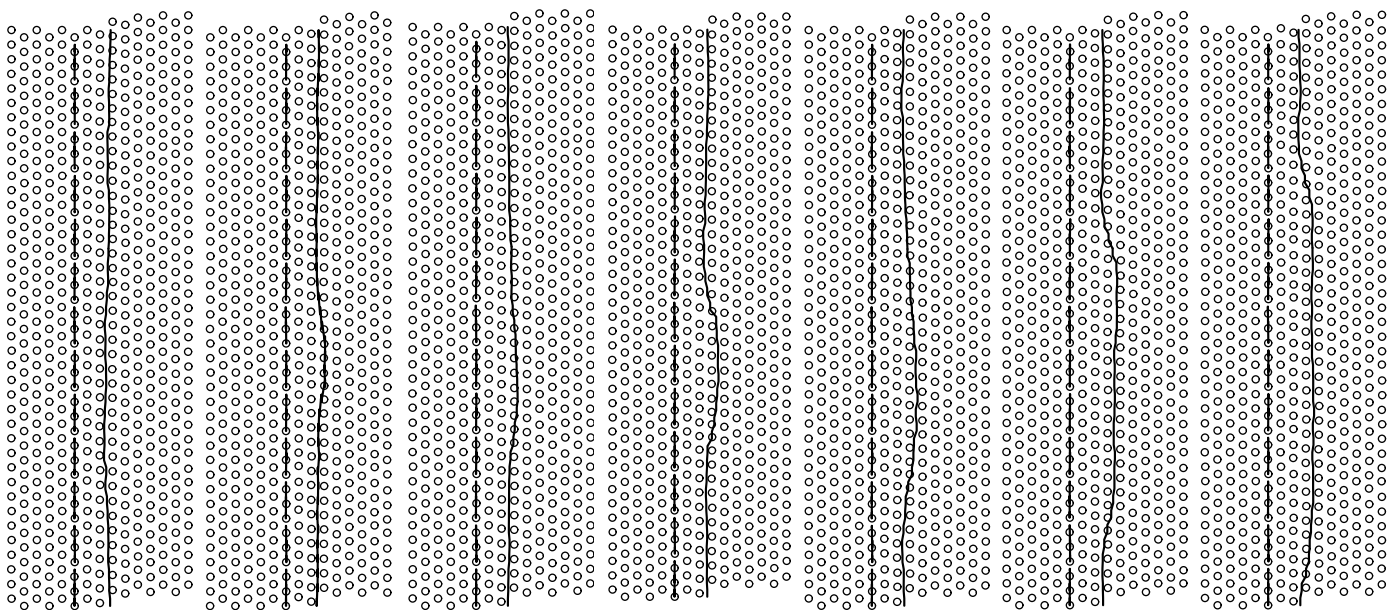
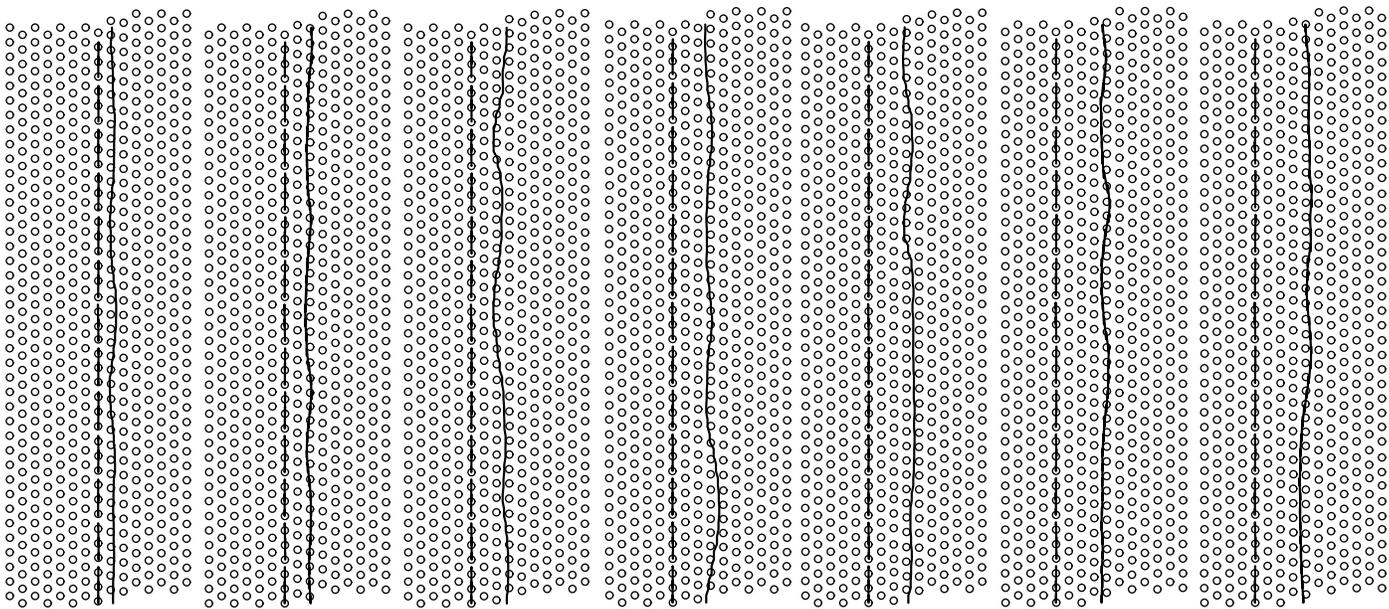
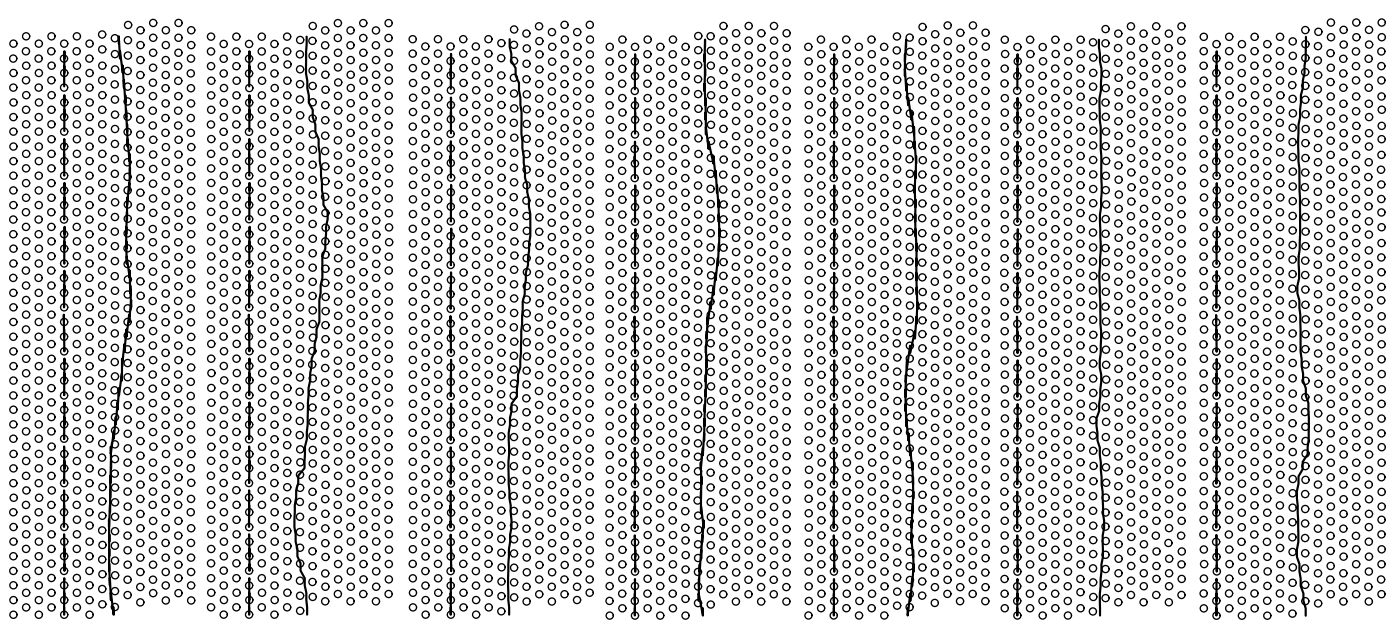
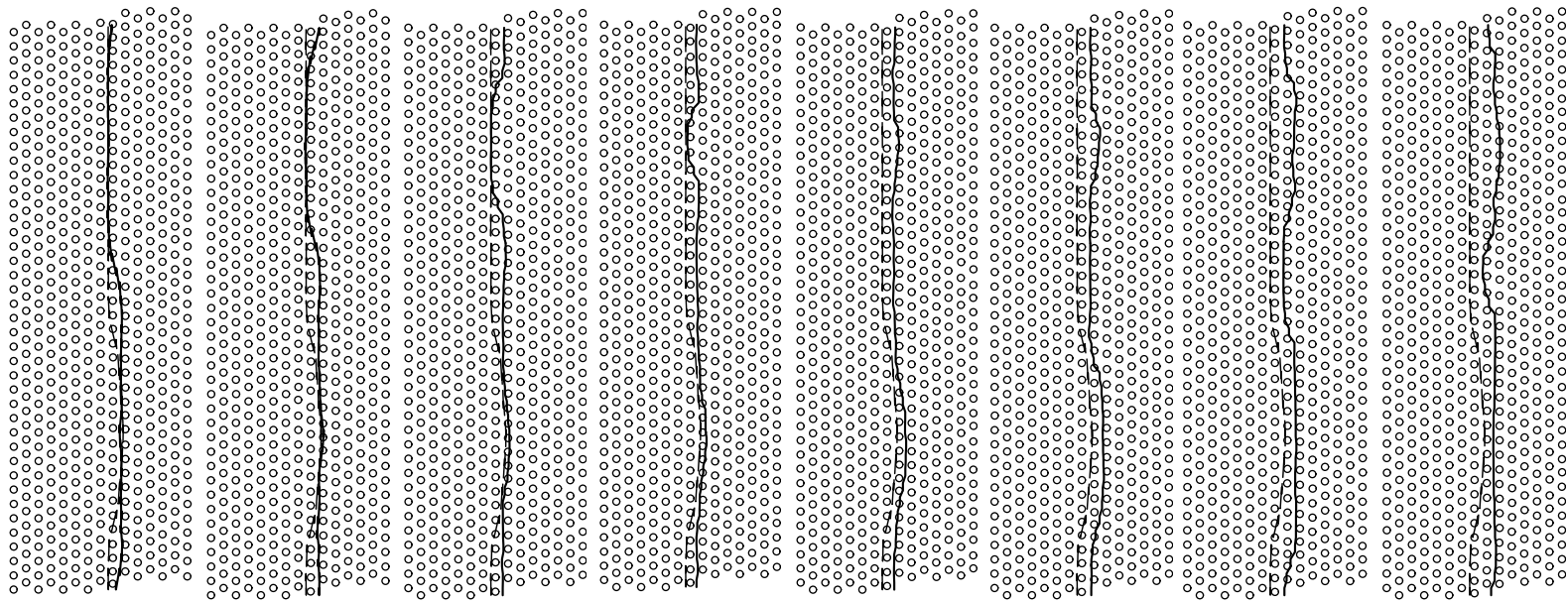
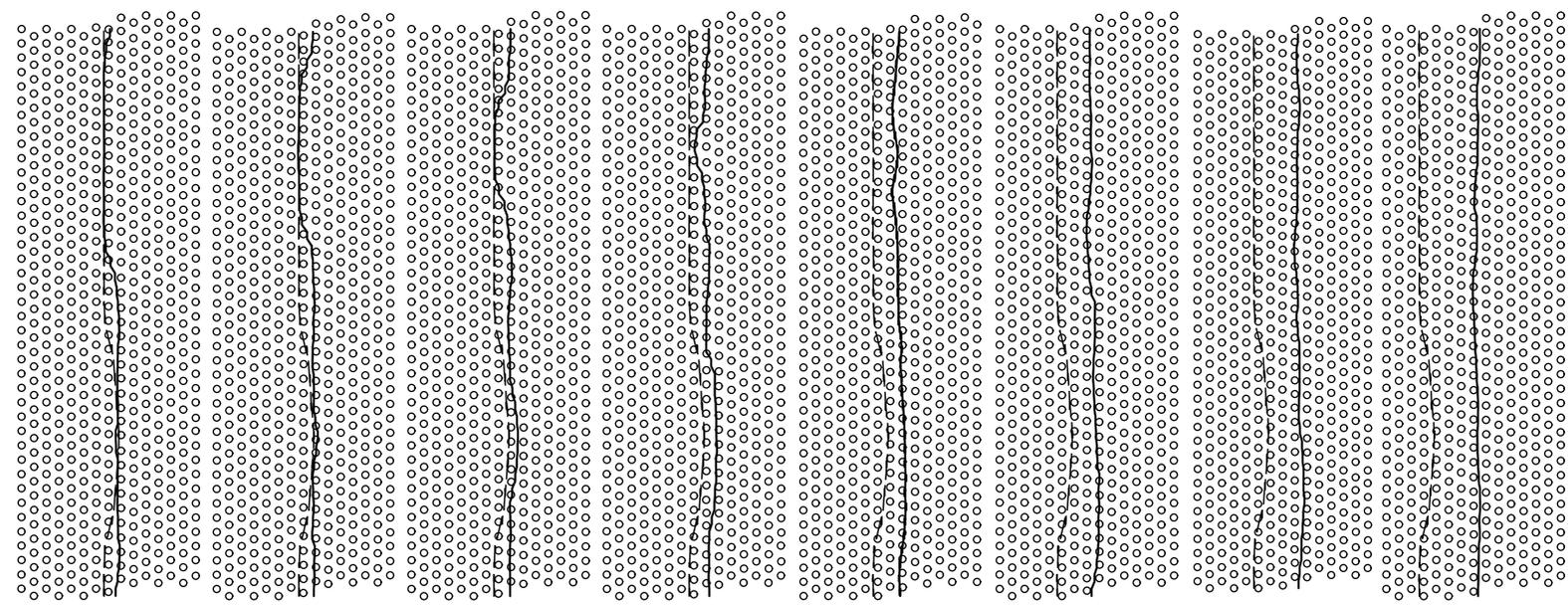
\end{document}